\documentclass[runningheads]{llncs}

\usepackage[T1]{fontenc}

\usepackage{graphicx} \graphicspath{ {images/} } 
\usepackage[numbers]{natbib}
\usepackage{amsmath, bbm}
\usepackage{microtype, booktabs, multicol, multirow}
\usepackage{subcaption}

\usepackage{float}
\usepackage{pgfplots}
\usepackage{adjustbox}
\usepackage{amssymb}
\usepackage{multirow}

\usepackage[pdftex, pagebackref=true, pdfpagelabels=true,
hypertexnames=true, plainpages=false, naturalnames=false,
bookmarks=true, bookmarksnumbered=true,
pdfpagemode=UseOutlines,]{hyperref}
\hypersetup{colorlinks, citecolor=[rgb]{0,0.7,0}, filecolor=[rgb]{0.6,0.0,0.0}, linkcolor=[rgb]{0.0,0.0,0.5}, urlcolor=[rgb]{0.6,0.0,0}}

\usepackage{color}

\begin{document}

\title{Autoadaptive Medical Segment Anything Model}

\author{Tyler Ward \and Meredith K. Owen \and O'Kira Coleman \\ Brian Noehren \and Abdullah-Al-Zubaer Imran}
\authorrunning{Ward et al.}
\institute{University of Kentucky, Lexington, KY 40506, USA\\
    \email{tyler.ward@uky.edu}}

\maketitle

\begin{abstract}
Medical image segmentation is a key task in the imaging workflow, influencing many image-based decisions. Traditional, fully-supervised segmentation models rely on large amounts of labeled training data, typically obtained through manual annotation, which can be an expensive, time-consuming, and error-prone process. This signals a need for accurate, automatic, and annotation-efficient methods of training these models. We propose \textit{ADA-SAM} (automated, domain-specific, and adaptive segment anything model), a novel multitask learning framework for medical image segmentation that leverages class activation maps from an auxiliary classifier to guide the predictions of the semi-supervised segmentation branch, which is based on the Segment Anything (SAM) framework. Additionally, our ADA-SAM model employs a novel gradient feedback mechanism to create a learnable connection between the segmentation and classification branches by using the segmentation gradients to guide and improve the classification predictions. We validate ADA-SAM on real-world clinical data collected during rehabilitation trials, and demonstrate that our proposed method outperforms both fully-supervised and semi-supervised baselines by double digits in limited label settings. Our code is available at: \url{https://github.com/tbwa233/ADA-SAM}.


\end{abstract}

\section{Introduction}
Muscle dysfunction following knee injuries and surgical interventions is a major challenge in orthopedic rehabilitation. For example, injuries such as patellar instability or anterior cruciate ligament tear often result in quadriceps weakness~\cite{ref_sahraoui}, altered knee biomechanics~\cite{ref_henderson}, and increased risk for long-term joint dysfunction~\cite{ref_gao}. Current clinical assessments of muscle recovery after injury often rely on manual strength testing and subjective patient-reported outcomes, with limited image-based evaluation methods available, indicating a need for automated, objective, and quantitative techniques. 

One of the key challenges in post-injury quadriceps rehabilitation is the imbalanced recovery of the vastus lateralis (VL) and vastus medialis (VM) muscles~\cite{ref_hosseini}. These two quadriceps muscles play a crucial role in stabilizing the knee joint, maintaining proper patellar tracking, and preventing recurrent instability. Weakness or delayed activation of the VM in particular is associated with poor knee control and increased stress on the ACL graft~\cite{ref_he}. Similarly, excessive lateral pull from the VL can contribute to maltracking of the patella, leading to long-term cartilage damage and post-traumatic osteoarthritis (PTOA) progression~\cite{ref_deuja}. Automated segmentation of VL and VM from medical imaging can provide novel ways to track muscle hypertrophy, atrophy, and symmetry during rehabilitation.

Medical image segmentation is a powerful method for enabling precise, reproducible, and efficient muscle delineation from imaging data. In particular, deep learning-based models, such as U-Net~\cite{ref_ronneberger} and its variants, have demonstrated state-of-the-art performance in segmenting musculoskeletal structures. These models have been widely applied to tasks such as muscle fiber tracking~\cite{ref_du}, lesion detection~\cite{ref_guo}, and volumetric analysis~\cite{ref_ironside}, and have shown promise for clinical decision support~\cite{ref_saad}. However, the adoption of deep learning for muscle segmentation in rehabilitation trials remains largely unexplored, in part due to the scarcity of large labeled datasets for supervised training.

Self-supervised learning (SSL) techniques leverage both labeled and unlabeled data, reducing the dependency on large manually labeled datasets while still achieving high segmentation accuracy~\cite{imran2020self}. By incorporating unlabeled data into the training process, SSL models can learn more robust feature representations and generalize better across diverse imaging datasets~\cite{ref_engelen, ref_haque}. This is particularly valuable for rehabilitation clinical studies, where imaging data is available but extensive manual annotations are infeasible.

Going beyond simple segmentation, multitask learning (MTL) frameworks that combine segmentation with auxiliary tasks, such as muscle classification~\cite{ref_zhou_a} or structural analysis~\cite{ref_zhang}, can further enhance performance. Recent work has shown that classification-guided segmentation, where class activation maps (CAMs) from a classification model provide spatial priors for segmentation, can improve accuracy and robustness in medical imaging applications~\cite{ref_ward}. In the muscle segmentation for large-scale clinical studies, such a multitask approach could help identify clinically relevant muscle regions, track muscle adaptation over time, and provide more interpretable segmentation outputs. We, therefore, propose a semi-supervised multitask learning framework for the segmentation of thigh magnetic resonance images (MRIs). Our contributions can be summarized as: 

\begin{itemize}
\item A novel segmentation model (ADA-SAM) that self-prompts,  provides self-feedback, and self-corrects;
\item An innovative qualitative evaluation of segmentation performance (SegEx) with both domain and large language model (LLM) experts;
\item A unique multiclass quadriceps muscles (VL and VM)-annotated dataset of 21 sequences collected from a Turbo Spin Echo MRI scanner;
\item Extensive experimentation demonstrates impressive performance even when trained on as few as 5 segmentation-labeled MRI slices.
\end{itemize}

\begin{figure}[t]
\centering
\includegraphics[width=\linewidth]{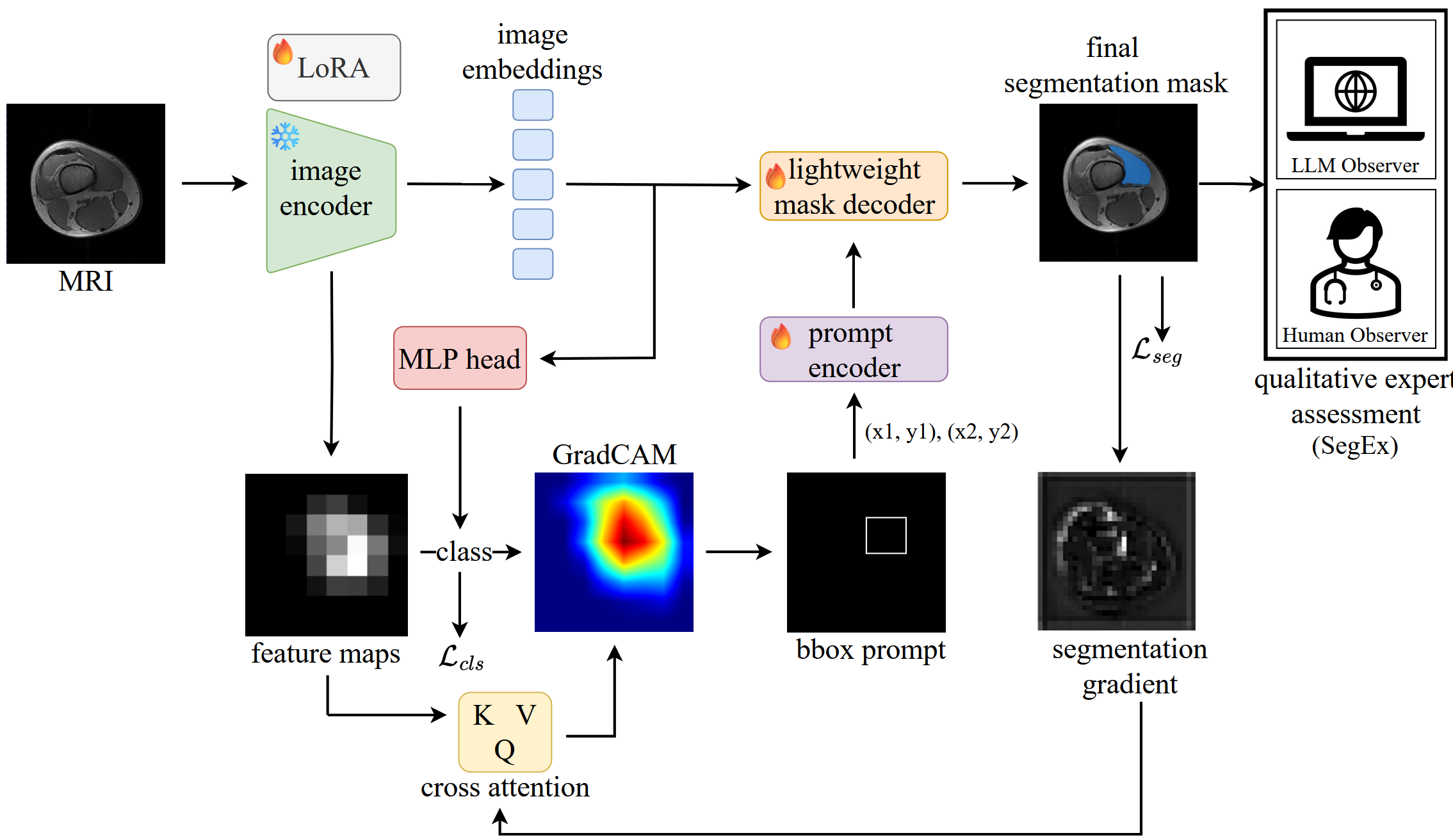}
\caption{Overview of the ADA-SAM framework. ADA-SAM is an autoadaptive, multitask learning model that utilizes a classification branch to enable automated prompt generation for the SAM-based segmentation branch. ADA-SAM utilizes a gradient feedback mechanism, where segmentation gradients refine classification representations, serving as task guidance for the segmentation. LoRA is applied to SAM’s transformer encoder for parameter-efficient domain adaptation.}
\label{fig:ADA-SAM}
\end{figure}

\section{Methods}

We propose a novel SAM-based segmentation model (ADA-SAM) without requiring any manual prompting and refinement. Additionally, an innovative, clinically interpretable assessment scheme (SegEx) is implemented to evaluate the segmentation performance. Fig.~\ref{fig:ADA-SAM} demonstrates the segmentation and evaluation components in our proposed approach.

\subsection{ADA-SAM}
To formulate the problem, we assume a data distribution $p(\mathit{X}, \mathit{Y})$ over $\mathit{D}$ where $\mathit{X} = \{x_1, x_2, \dots, x_N\}$ is a set of MRI image slices and $\mathit{Y} = \{y_1, y_2, \dots, y_N\}$ is the set of corresponding ground truth segmentation masks. A network $\mathit{G}_{\phi}$ is created with parameters $\phi$ such that $\mathit{G}_{\phi}(\mathit{X}) \rightarrow \mathit{Y}$. To aid in segmentation, an auxiliary classification task is manufactured by creating a set of class labels $\mathit{C}$ such that $\mathit{C} = \{\max(\mathit{y}_i)\}$ where $\mathit{y}_i = \{0, 1, 2, 3\}^{256\times256}$ for $i = 1, \ldots, N$, with $N$ being the number of slices in $D$.

\noindent\textbf{Automated Prompt Generation:}  
The true power of any SAM-based segmentation model lies in prompt generation. In ADA-SAM, we use GradCAMs obtained from the classification branch to initialize the segmentation branch. These GradCAMs highlight the most discriminative image regions, forming the basis for automated prompt generation.
Given an input image \( x \), the classification branch predicts a class \( c = \arg\max p \), where \( p \in [0,1]^C \) represents class probabilities. To identify key regions, we generate a class activation map \( M_{\text{CAM}} \in \mathbb{R}^{H \times W} \), where pixel intensities reflect their importance in classification.
Applying a threshold \( \tau \) isolates the most salient areas, and the bounding box enclosing these activated pixels defines the region of interest (ROI). The bounding box coordinates \( (x_{\min}, y_{\min}, x_{\max}, y_{\max}) \) serve as the prompt for SAM. SAM's image encoder extracts high-level features \( F(x) \), while the prompt encoder processes the bounding box into a learned representation \( P(B) \). The mask decoder then predicts the segmentation mask.

\noindent\textbf{Model and Data-Efficient Segmentation Prediction:}
In the ADA-SAM framework, the segmentation process begins by extracting feature representations from MRI slices using SAM’s image encoder. Given an input slice \( x \in \mathbb{R}^{H \times W} \), the encoder produces a feature representation: \( F(x) \in \mathbb{R}^{d \times h \times w} \), where \( d \) is the feature dimension and \( h \times w \) is the spatial resolution.
Fine-tuning SAM is computationally expensive and prone to overfitting due to its large number of parameters. To address this, we use low-rank adaptation (LoRA), a parameter-efficient fine-tuning technique. LoRA introduces trainable low-rank matrices \( A \) and \( B \) with rank \( r \) to modify SAM’s transformer attention layers. Setting \( r = 8 \) constrains adaptation while maintaining expressive power, enabling domain-specific learning with minimal overhead.
To balance efficiency and learning, segmentation follows a semi-supervised approach where only a subset of slices are labeled. Selection prioritizes class representation by ensuring at least one sample per class, with remaining samples chosen randomly to meet the target number.
Once labeled samples are determined, segmentation is guided by prompts from the classification branch. Given an encoded feature representation \( F(x) \) and an encoded prompt \( P(B) \), the segmentation decoder predicts the final segmentation mask.

\noindent\textbf{Mask Refinement with Self-Feedback:}  
The self-feedback mechanism in ADA-SAM enables iterative refinement by linking classification and segmentation branches. This bidirectional feedback ensures classification-based activations enhance segmentation prompts while segmentation-derived gradients refine classification features.
First, the classification branch generates a class prediction, producing a class activation map \( M_{\text{CAM}} \), which is thresholded to extract a bounding box \( B \). The prompt encoder processes \( B \) into a segmentation prompt \( P(B) \), which, along with encoded features, guides segmentation mask generation.
Segmentation gradients propagate back to refine shared feature representations. The updated features \( F'(x) \) improve classification predictions, reinforcing consistency between classification and segmentation. This self-feedback mechanism enhances segmentation accuracy while maintaining classification robustness.

\noindent\textbf{Training Objectives:}
During training, we employ a multitask loss function to jointly optimize the classification and segmentation tasks within the ADA-SAM framework. This is done by combining the classification and segmentation losses along with a weighting factor. For classification, we minimize the Focal loss~\cite{lin2017focal} due to the high class imbalance in our training set. The Focal loss is defined as:
\begin{equation}
    \mathcal{L}_{\text{cls}} = - \sum_{c=1}^{C} \alpha_c (1 - p_{t,c})^{\gamma} k_c \log(p_{t,c}),
\end{equation}
where $C$ is the number of classes, $k_c$ is the one-hot encoded ground truth for class $c$, $p_{\text{t, c}}$ is the predicted probability of class $c$, and $\gamma$ is the class-specific weighting factor. For segmentation, we minimize the Dice loss as:
\begin{equation}
\mathcal{L}_{\text{seg}} = 1 - \frac{2 \sum_{i} \hat{y}_i y_i}{\sum_{i} \hat{y}_i + \sum_{i} y_i},
\end{equation}
where $\hat{y}$ is the predicted segmentation. Therefore, the overall objective for ADA-SAM becomes
$\mathcal{L}_{\text{mtl}} = \mathcal{L}_{cls} + \gamma\mathcal{L}_{
\text{seg}}$,
where $\gamma$ is the weighting factor that balances the two losses.

\subsection{SegEx: Expert-driven Segmentation Assessment}
To evaluate the clinical applicability and reliability of ADA-SAM in segmenting the VL and VM muscles, we introduce an innovative expert assessment (SegEx). SegEx is conducted by setting up a visual Turing test employing expert observers --- human or large language model (LLM) observers. The goal is to assess the quality of the ADA-SAM generated masks for their clinical utility. First, consulting with domain expert collaborators, we select a set of clinically-relevant evaluation criteria $E = \{e_1, e_2, \dots, e_k\}$. As defined in Table~\ref{tab:criteria}, the selected criteria focus on mask quality, size difference, whether any correction is needed, and diagnostic confidence. Each of the criteria may have different scales; they are standardized ($\hat{e}_i$) to be in a common range $[r_{min}$, $r_{max}]$. Any binary decision (0/1) criteria are multiplied by the maximum $r_{max}$. The final evaluation score is then obtained by calculating the average across all the criteria, i.e., $E_{avg} = \frac{1}{k}\sum^k_1 \hat{e}$.   

\noindent \textbf{Human observer:} SegEx is performed by selecting two 3D MRI test cases with their corresponding 3D segmentation masks (ground truth $Y$ and prediction $\hat{Y}$). To avoid bias, $Y$ and $\hat{Y}$ are randomly mixed without indicating their sources. Two human observers of different levels of expertise participated in this study, which ensured an unbiased evaluation. The assessment is performed without any inter-observer communication and in a randomized order to minimize further bias in the evaluation. 

\noindent \textbf{LLM observer:} SegEx for LLM observers is performed with the multimodal LLM, GPT-4o. Unlike the human observers, the LLMs are presented with only the segmentation masks ($Y$, $\hat{Y}$) to avoid any privacy violation. Additionally, ratings for the ED metric were not provided by the LLM, as this metric relies on private bioinformatic data unavailable to the models.

\begin{table}[t]
    \centering
    \caption{For scoring the human evaluations, we follow an approach similar to~\cite{ref_chen_b}, where a 1-4 scale is used, with 1 being the best-case and 4 being the worst. The one exception to this 1-4 scale is the CN metric, which is a binary decision: no (0) or yes (1).}
    \label{tab:criteria}
    \resizebox{\linewidth}{!}{
    \begin{tabular}{lc cc p{8cm}}
        \toprule
        \textbf{Criteria} & \phantom{a} & \textbf {Scale} & \phantom{a} & \textbf{Description}\\
        \midrule
        General mask quality (MQ) && 1-4 && What is the quality of the segmentation mask? Are there any gaps, rough edges, etc.\\
        \midrule
        Mask boundaries (MB) && 1-4 && How well do the masks encompass the muscles? Are there over-segmentations, under-segmentations, etc.?\\
        \midrule
        Correction needed (CN) && 0/1 && Would a clinician need to modify the mask?\\
        \midrule
        Size difference (SD) && 1-4  && How closely do the predicted masks match the size of the ground truth masks?\\
        \midrule
        Diagnostic confidence (DC) && 1-4 && Confidence that the masks could be used for diagnostic volume assessment.\\
        \bottomrule
    \end{tabular}
    }
\end{table}

\section{Experimental Evaluation}
\subsection{Implementation Details}
\textbf{Data:}
We validated our proposed ADA-SAM method on a dataset obtained from individuals with patellar instability \cite{ref_brightwell}. This is a multi-class dataset containing segmentation information for two quadriceps muscles: VL and VM. The data was collected from a Turbo Spin Echo MRI sequence in 65 slice stacks with a field of view of 192$\times$192 mm, an acquisition matrix of 256$\times$256, 6 mm slice thickness, and a voxel size of 0.75$\times$0.75$\times$6 mm. Originally, the dataset contained scans from 21 patients, which we split into train, validation, and test sets using a 15/1/5 split containing 975/65/325 slices, respectively.

\textbf{Baselines:}
For baseline comparisons, we compare ADA-SAM against both fully supervised single-stage models~\cite{ref_ronneberger, ref_zhou_b, ref_isensee, ref_chen_a, ref_cao} and semi-supervised multitask learning frameworks~\cite{ref_ward}.

\textbf{Training:}
Each non-SAM-based baseline model was trained for 25 epochs. Following ~\cite{ref_ward}, which demonstrated that SAM models could outperform more traditional deep learning baselines with fewer epochs, we train ADA-SAM with 10 epochs only.

\textbf{Machine Configuration:} The models are trained on a \emph{Intel (R) Xeon (R) w7-2475X, 2600MHz} machine with a dual \emph{NVIDIA A4000X2} GPUs (32GB).

\textbf{Evaluation:}
For quantitative evaluation, we used the Dice similarity coefficient (DSC) as a measure of similarity. We report the results after human and LLM expert evaluation for qualitative evaluation. For this qualitative evaluation, three full 3D MRI scans from the test set were given to the evaluators, and each slice in the scan was evaluated. The results reported are the averages and standard deviations of the slices from the three scans.

\begin{table}[t]
    \centering
    \caption{Quantitative comparison of ADA-SAM demonstrates the superiority in segmenting VL and VM from MRIs in a limited annotated setting: Avg(stdev) Dice scores are reported after 5 runs of each of the models. The best and second-best results are \textbf{bolded} and \underline{underlined}, respectively.}
    \label{tab:combined}
    \resizebox{\linewidth}{!}{
    \small
    \begin{tabular}{l c c c c c c c c}
        \toprule
        & \multicolumn{8}{c}{Fully Supervised Models} \\
        \cmidrule{2-9}
        Class & UNet & UNet++ & Trans-UNet & nnU-Net & Swin-Unet & MedSAM & SAM-Mix & ADA-SAM \\
        \midrule
        VM  & 0.76(0.017) & 0.79(0.123) & 0.74(0.001) & \underline{0.83(0.139)} & \textbf{0.88(0.014)} & 0.71(0.028) & 0.82(0.002) & 0.82(0.006) \\
        VL & 0.76(0.010) & 0.80(0.165) & 0.74(0.001) & \textbf{0.81(0.127)} & 0.80(0.013) & 0.69(0.068)  & 0.80(0.009) & 0.80(0.002) \\
        Overall & 0.76(0.014) & 0.80(0.144) & 0.74(0.001) & \underline{0.82(0.138)} & \textbf{0.84(0.014)} & 0.70(0.048) & 0.81(0.006) & 0.81(0.004) \\
    \end{tabular}
    }
    \resizebox{\linewidth}{!}{
    \begin{tabular}{c c c c  c c c c c}
        \toprule
        && \multicolumn{7}{c}{Semi-Supervised Models} \\
        \cmidrule{3-9}
        \#Slices & Class & UMTL & UMTL++ & nnUMTL & SAM-Mix & SAM-PP & ADA-SAM* & ADA-SAM \\
        \midrule
        \multirow{3}{*}{0} 
        & VL  & 0.16(0.114)  & 0.19(0.089)  & 0.26(0.012) & 0.61(0.037) & 0.46(0.061) & \underline{0.85(0.023)} & \textbf{0.88(0.016)}  \\
        & VM  & 0.21(0.160)  & 0.22(0.068)  & 0.30(0.005) & 0.63(0.022)   & 0.55(0.024) & \underline{0.87(0.017)} &  \textbf{0.90(0.009)}\\
        & Overall  & 0.19(0.082)  & 0.21(0.079)  & 0.28(0.114) & 0.62(0.035)   & 0.51(0.043) & \underline{0.86(0.025)} &  \textbf{0.89(0.018)} \\
        \midrule
        \multirow{3}{*}{5} 
        & VL  & 0.30(0.031)  & 0.43(0.016)  & 0.48(0.093)  & 0.77(0.010)   & 0.60(0.048) & \underline{0.88(0.015)} &  \textbf{0.91(0.002)} \\
        & VM  & 0.40(0.006)  & 0.45(0.024)  & 0.51(0.079)  & 0.77(0.001)   & 0.62(0.023) & \underline{0.89(0.016)} &  \textbf{0.92(0.000)} \\
        & Overall  & 0.35(0.035) & 0.44(0.025)  & 0.50(0.086)  & 0.77(0.021)   & 0.61(0.036) & \underline{0.89(0.016)} &  \textbf{0.92(0.001)}  \\
        \midrule
        \multirow{3}{*}{50} 
        & VL  & 0.54(0.138)  & 0.62(0.095)  & 0.69(0.052)   & 0.80(0.047)   & 0.64(0.054) & \underline{0.89(0.011)} &  \textbf{0.94(0.019)}\\
        & VM  & 0.61(0.053)  & 0.61(0.036)  & 0.70(0.030)   & 0.82(0.048)    & 0.68(0.035) & \underline{0.92(0.012)} &  \textbf{0.95(0.017)}\\
        & Overall  & 0.58(0.096)  & 0.62(0.066)  & 0.70(0.041)   & 0.81(0.048)  & 0.66(0.035) & \underline{0.91(0.012)} &  \textbf{0.95(0.018)}\\
        \midrule
        \multirow{3}{*}{100} 
        & VL  & 0.62(0.084)  & 0.76(0.019)  & 0.77(0.051)  & 0.86(0.023)  & 0.68(0.026) & \underline{0.89(0.000)} &  \textbf{0.91(0.018)}  \\
        & VM  & 0.68(0.065)  & 0.75(0.002)  & 0.76(0.037)  & 0.86(0.004)   & 0.70(0.019) & \underline{0.91(0.001)} &  \textbf{0.94(0.003)}  \\
        & Overall  & 0.65(0.075)  & 0.76(0.016)  & 0.77(0.044)  & 0.86(0.024) & 0.69(0.028) & \underline{0.90(0.001)} & \textbf{0.92(0.016)}  \\
        \bottomrule
    \end{tabular}
    }
    \footnotesize{\textit{*Without cross-attention.}}
\end{table}

\begin{figure}[t]
    \centering
    \resizebox{0.95\linewidth}{!}
    {
    \begin{tabular}{c c c c c}
     {\large nnU-Net} & {\large Swin-Unet} & \multicolumn{3}{c}{\large ADA-SAM (L$\rightarrow$R): 5, 50, and 100 slices}\\
     \smallskip
     
     \includegraphics[width=0.19\linewidth]{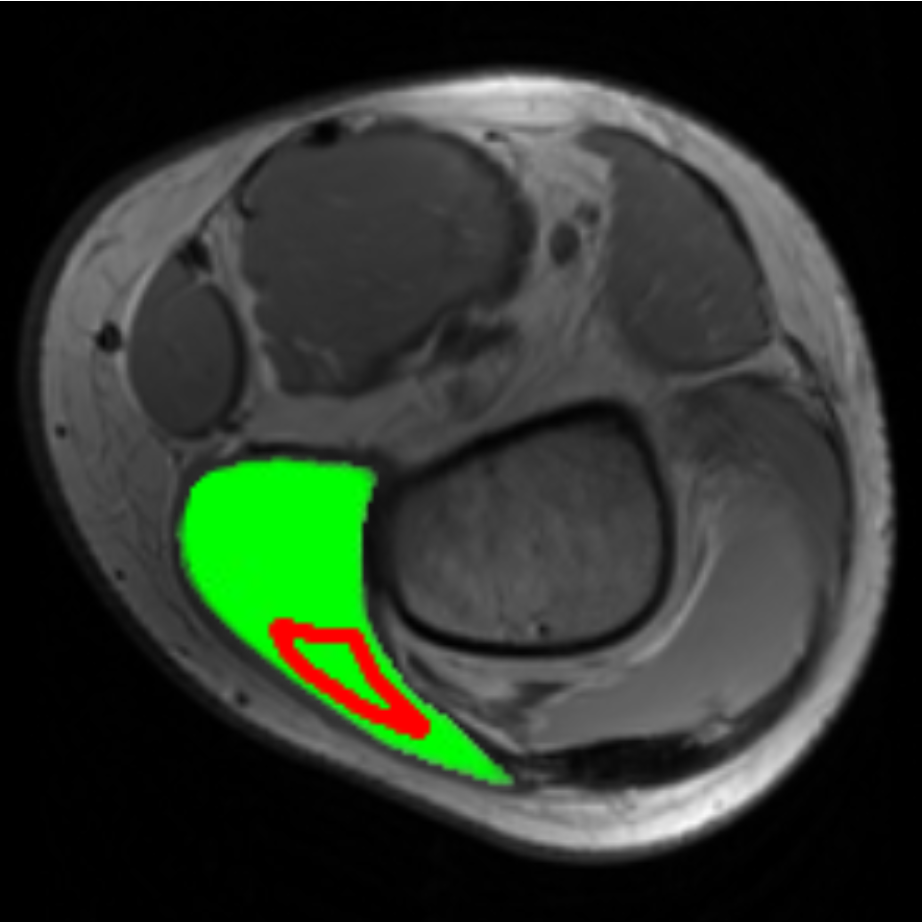}
     &
     \includegraphics[width=0.19\linewidth]{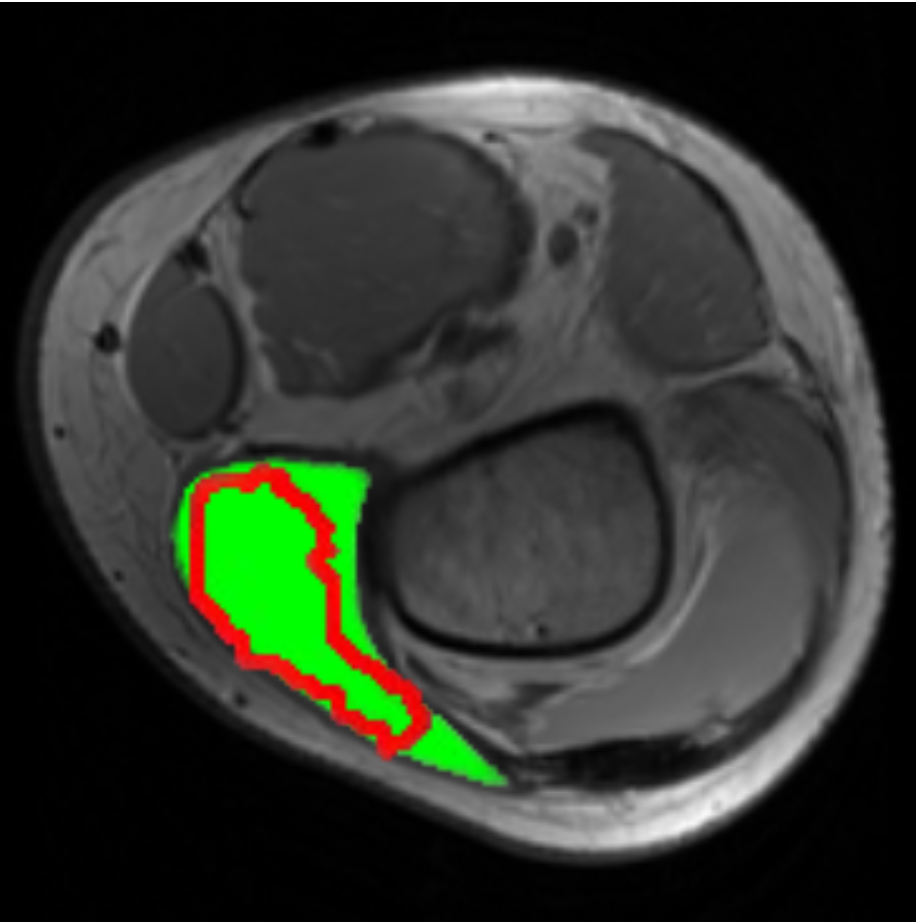}
     &
     \includegraphics[width=0.19\linewidth]{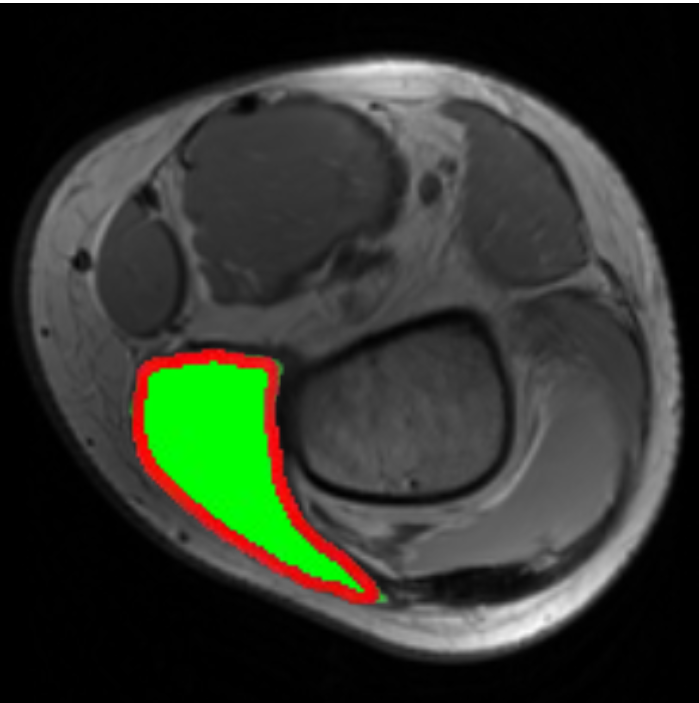}
     &
     \includegraphics[width=0.19\linewidth]{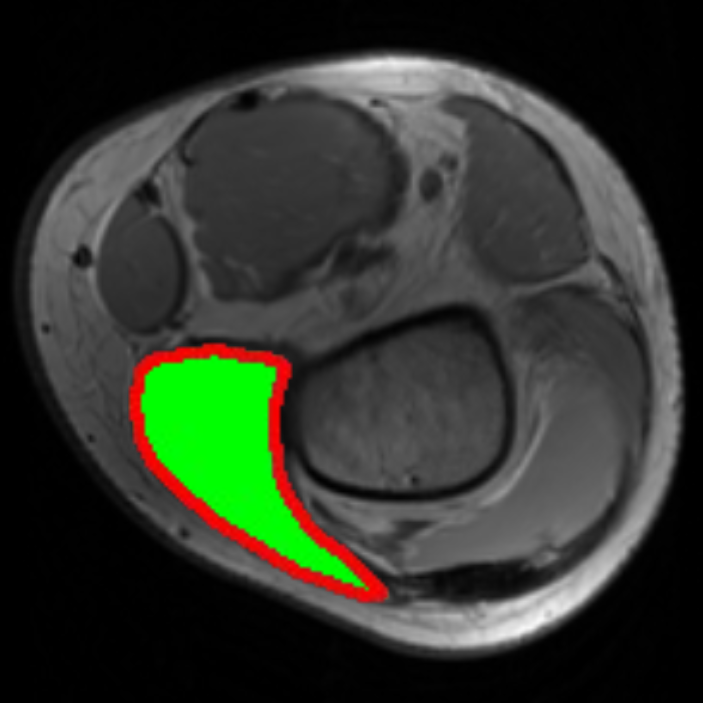}
     &
     \includegraphics[width=0.19\linewidth]{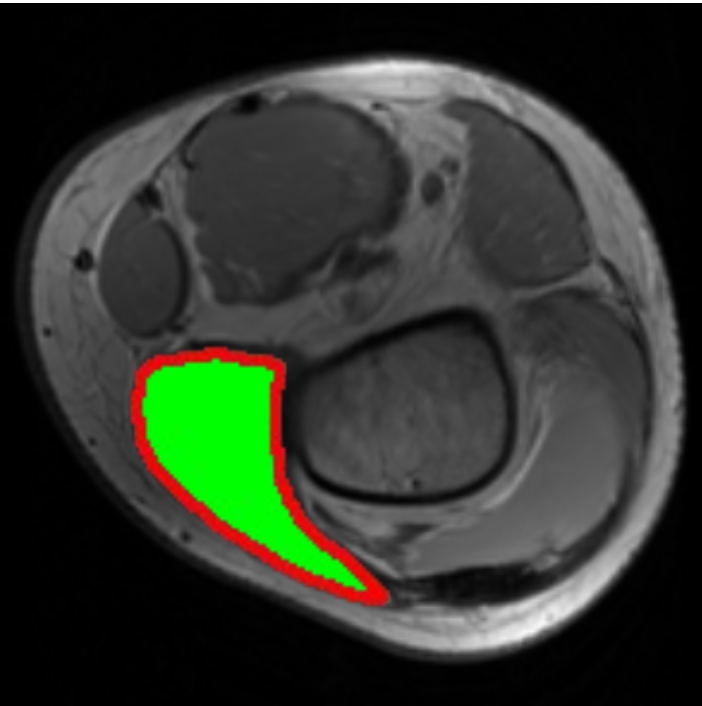}
     \\
     \includegraphics[width=0.19\linewidth]{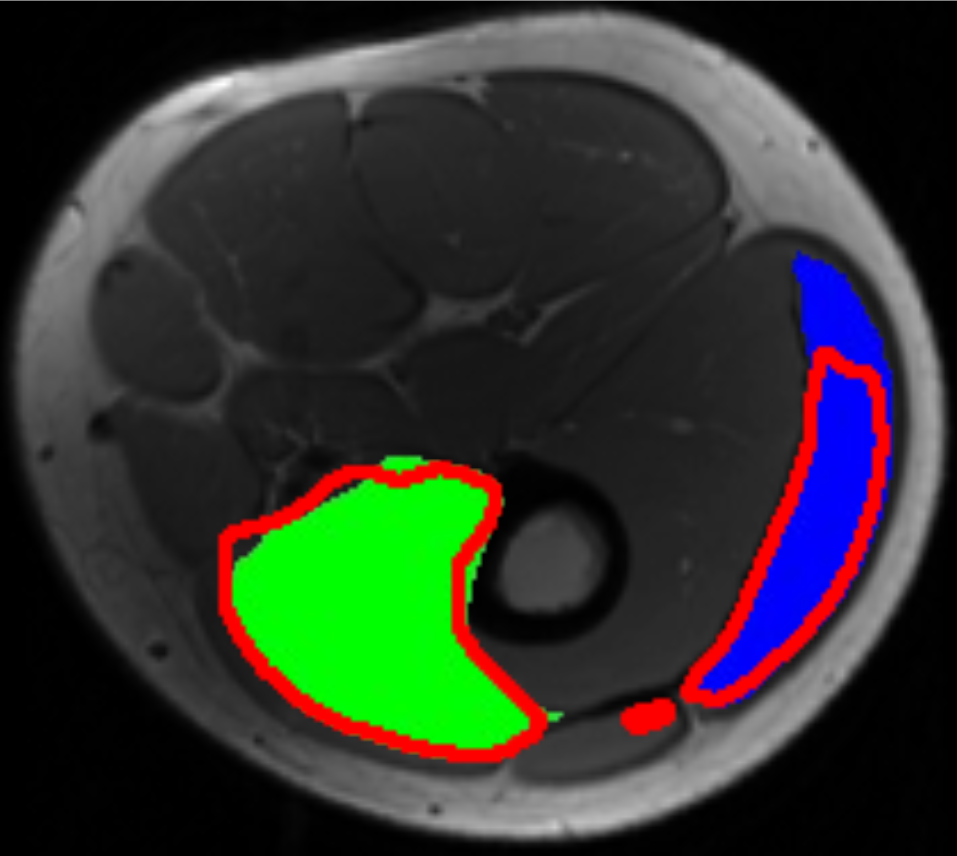}
     &
     \includegraphics[width=0.19\linewidth]{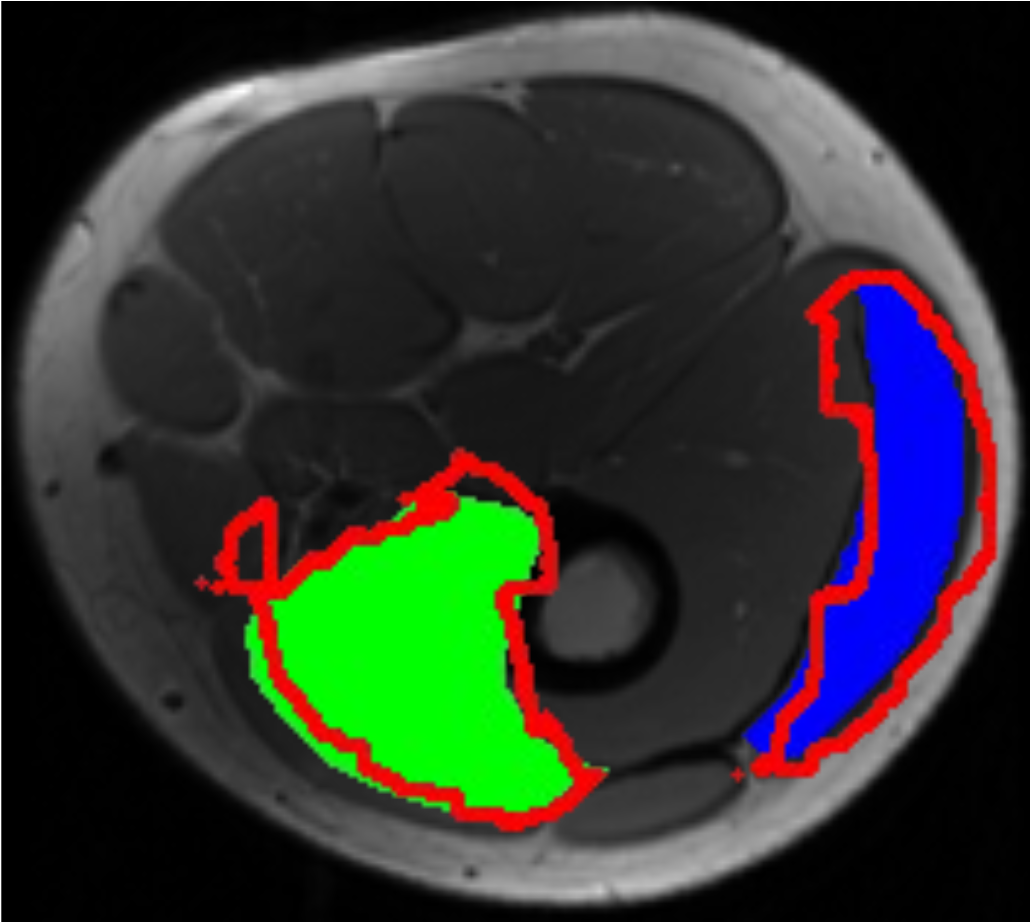}
     &
     \includegraphics[width=0.19\linewidth]{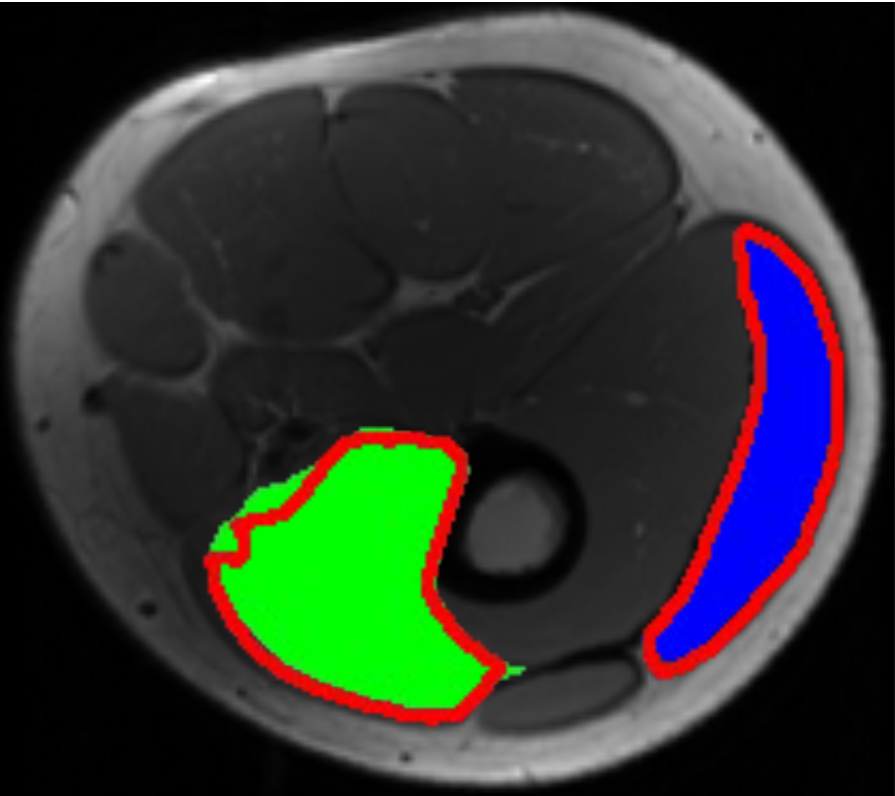}
     &
     \includegraphics[width=0.19\linewidth]{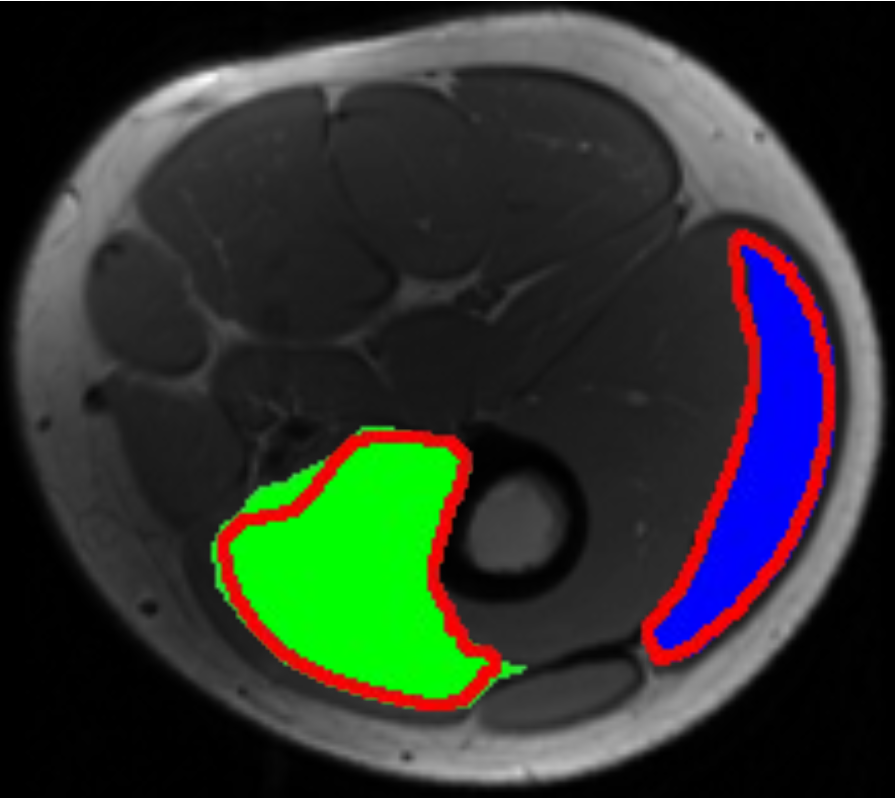}
     &
     \includegraphics[width=0.19\linewidth]{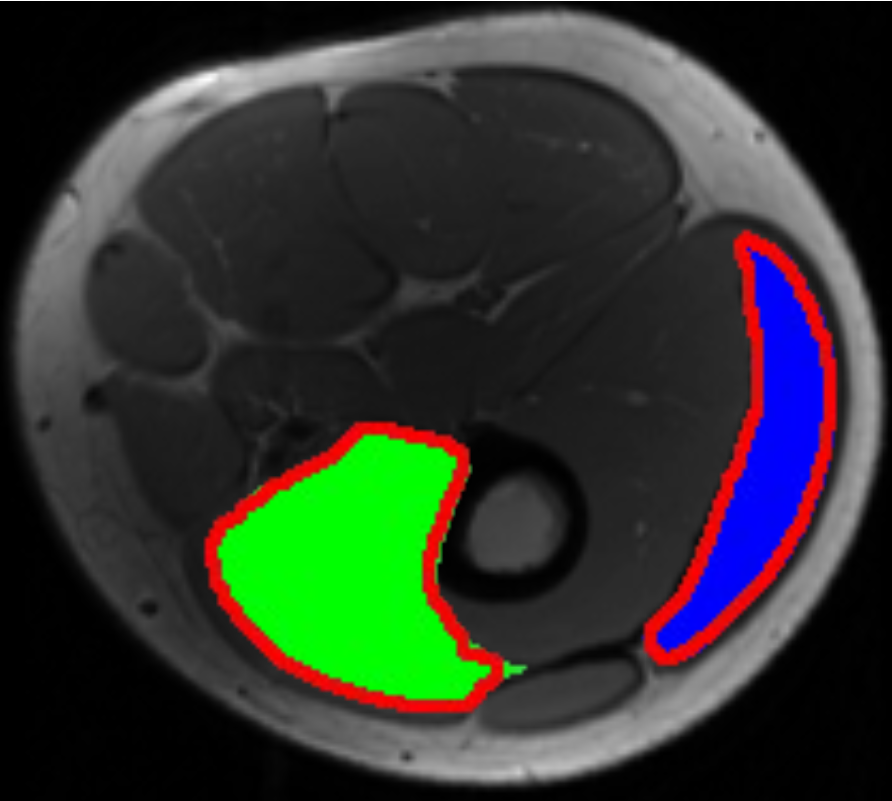}
    \end{tabular}
    }
    \caption{Qualitative comparison demonstrates the superiority of semi-supervised ADA-SAM over the best-performing fully supervised model in segmenting VL and VM
    Color code: Green - ground truth VM, Blue - ground truth VL, Red - predicted contour.
    }
    \label{fig:masks}
\end{figure}

\subsection{Results and Discussion}
\textbf{Quantitative Evaluation:} Table~\ref{tab:combined} reports ADA-SAM's fully-supervised performance compared to other fully-supervised models, as well as its semi-supervised performance compared against other semi-supervised multitasking models. Given that ADA-SAM is designed as a semi-supervised model, allowing it to leverage both labeled and unlabeled data, leading to improved regularization and generalizability, it is expected that ADA-SAM's fully supervised setting is competitive but does not surpass the baseline models. The power of ADA-SAM is clearly evident in the semi-supervised setting, with it outperforming other semi-supervised multitask models across different labeled data settings. With just 5 labeled slices, ADA-SAM achieves a DSC of 0.91, surpassing SAM-Mix (DSC 0.77) and significantly outperforming the third-best method by 31\% (DSC). This trend continues at 50 and 100 labeled slices, where ADA-SAM consistently surpasses SAM-Mix~\cite{ref_ward} and other approaches in DSC. 

Fig.~\ref{fig:masks} demonstrates a qualitative comparison between the semi-supervised ADA-SAM and the best-performing fully-supervised methods (nnU-Net and Swin-Unet). For binary cases, where only one muscle is present in the slice, ADA-SAM addresses the issue of under-segmentation that is present for nnU-Net and Swin-Unet. For slices containing both muscles, ADA-SAM demonstrates improvements in fixing under-segmentations in the case of nnU-Net, and fixing over-segmentations in the case of Swin-Unet.

\noindent \textbf{Qualitative Expert Assessment:} Table~\ref{tab:assessment} reports the results from the qualitative evaluations, showing that while ADA-SAM-generated masks were distinguishable from ground truth, they were rated favorably, particularly for VM segmentation. VM's higher ratings likely stem from its smaller size and simpler shape in MRI scans. Notably, evaluators ranked ADA-SAM’s VM segmentations as having better DSC scores than the ground truth in some cases, suggesting its potential for enhancing subsequent clinical decision making.

\begin{table}[t]
    \centering
    \caption{Qualitative assessment (SegEx) of the ADA-SAM predicted (P) VL and VM muscles compared to the ground truth (GT). For each observer, avg(stdev) scores are reported over the entire test set. 
    }
    \label{tab:assessment}
    \resizebox{0.83\linewidth}{!}{
    \begin{tabular}{lc cc c cc cc c}
        \toprule
        Muscle & \phantom{a} &
        Source & \phantom{a} &
        DSC & Observer 1 & \phantom{a} & Observer 2 & \phantom{a} & LLM Observer \\
        \midrule
        VM && GT  && --- & 1.44(0.526) && 1.54(0.532) && 2.82(0.329)\\
        VM && P   && 0.97(0.080) & 1.48(0.603) && 1.42(0.607) && 2.77(0.341)\\
        VL && GT  && --- & 1.24(0.448) && 1.36(0.545) && 2.76(0.354)\\
        VL && P && 0.85(0.242) & 2.39(0.616) && 1.98(0.679) && 2.71(0.360)\\
        \bottomrule
    \end{tabular}
    }
\end{table}

\noindent \textbf{Ablation Experiments:}
We experiment with four different ranks to determine the optimal LoRA settings for ADA-SAM (2, 4, 6, 8). We found that a rank of 8 performs the best, achieving a DSC of 0.91 compared to 0.84, 0.85, and 0.86 for ranks 2, 4, and 6, respectively. Setting $r=8$ reduces the number of parameters to 26M compared to SAM's original count of 91M for the $ViT_B$ checkpoint. This also results in a large reduction of parameters from other existing SAM-based semi-supervised multitasking models~\cite{ref_ward}, requiring an external model for prompt generation. Furthermore, ADA-SAM has faster inference than SAM-Mix, taking approximately 132ms per image (vs 150ms), including I/O.

\section{Conclusions}
We have presented a novel semi-supervised, multi-task learning approach to automatically adapt (self-prompting and self-refining) SAM (ADA-SAM). 

This annotation-efficient (a small number of labeled slices) task guidance enables accurate segmentation of the vastus lateralis and vastus medialis muscles from MRI scans of the quadricep region, requiring no manual/semi-automated prompting. Experimental evaluations reveal that ADA-SAM outperforms baseline and state-of-the-art semi-supervised multi-task and SAM-based as well as generic fully-supervised segmentation models. In addition, our innovative expert assessment (SegEx) highlights the potential of clinically adopting ADA-SAM in analyzing injured muscles. Our ongoing work explores validating ADA-SAM on multimodal data by incorporating more robust human-in-the-loop feedback.

\bibliographystyle{splncsnat}
\bibliography{references}

\end{document}